# UNDERSTANDING THE PERCEPTION OF COVID-19 POLICIES BY MINING A MULTILANGUAGE TWITTER DATASET

**Christian E. Lopez**[*1,2], **Malolan Vasu**[1], **Caleb Gallemore**[3]

[1]Computer Science Department, Lafayette College, Easton, PA 18042
[2] Mechanical Engineering Department, Lafayette College, Easton, PA 18042
[3]International Affairs Program, Lafayette College, Easton, PA 18042

**ABSTRACT**

*The objective of this work is to explore popular discourse about the COVID-19 pandemic and policies implemented to manage it. Using Natural Language Processing, Text Mining, and Network Analysis to analyze corpus of tweets that relate to the COVID-19 pandemic, we identify common responses to the pandemic and how these responses differ across time. Moreover, insights as to how information and misinformation were transmitted via Twitter, starting at the early stages of this pandemic, are presented. Finally, this work introduces a dataset of tweets collected from all over the world, in multiple languages, dating back to January 22nd, when the total cases of reported COVID-19 were below 600 worldwide. The insights presented in this work could help inform decision makers in the face of future pandemics, and the dataset introduced can be used to acquire valuable knowledge to help mitigate the COVID-19 pandemic.*

Link for dataset:
https://github.com/lopezbec/COVID19_Tweets_Dataset

**INTRODUCTION**

The Coronavirus Disease of 2019, known as COVID-19, is a rapidly spreading disease caused by the Severe Acute Respiratory Syndrome Coronavirus 2 (SARS-CoV2). The COVID-19 is now considered a pandemic that has affected countries in all inhabited continents. Since the first cases of COVID-19 reported in Wuhan, China, in December 2019, the number of fatalities worldwide has increased rapidly. Due to its high infection and death rate, governments have implemented a wide range of policies aimed at mitigating the spread of this virus and its impact. Such actions began with the Chinese government order to quarantine Wuhan on January 23rd, 2020, to, most recently, multiple countries declaring state of emergency and implementing strict quarantine and social distancing measurements (e.g., US, Italy, Argentina, Spain).

Most government leaders have implemented measures to incentivize, and in some cases enforce, "social distancing" to reduce the spread of COVID-19. These measures have resulted in the cancelled entertainment events, closures of schools and colleges, and businesses reducing hours of operation, implement telecommuting, or close altogether. There is no doubt the pandemic and the measures set in place to mitigate it have and will continue to drastically impact the lives of millions. As this pandemic and the responses to it are unprecedented, however, we are likely to be surprised by how people respond.

Since the early stages of the disease, people have expressed their opinion and shared information, as well as misinformation, about it via social media platforms, such as Twitter. As COVID-19 spreads to other countries and governments try to mitigate its impact by implementing counter measures, people have also used social media platforms to express their opinion about the measures themselves, the leaders implementing them, and the ways their lives are changing. The use of social media, such as Twitter, as platforms to express opinions and share information about COVID-19, will only continue to grow, precisely because of the "social distancing" measures set in place to mitigate it.

Policymakers could mine this social media data to explore popular discourse about the pandemic and the measures set in place to mitigate it. We plan to analyze a corpus of tweets that relate to COVID-19 with the objective of identifying common responses to the pandemic and how these responses differ across time, countries, and policies. Moreover, insights as to how information and misinformation about this pandemic and the polices are transmitted are presented. Finally, we introduce and share with the research community a dataset of tweets collected from all over the world, in multiple languages, dating back to January 22nd when the total cases of reported COVID-19 were below 600 worldwide. Here, we describe and present descriptive

*Corresponding Author. 569 Rockwell Integrated Science Center, Lafayette College, Easton, PA 18042, lopezbec@lafayette.edu*

statistics of this dataset, and explain our data collection methods and intended analyses.

**DATASET DESCRIPTION**

The dataset presented is being continuously collected using the Twitter API. The dataset presented here (v1) covers 22 January to 13 March 2020 and contains 6,468,526 tweets. The keywords used for search tweets are: *virus* and *coronavirus* since 22 January, *ncov19* and *ncov2019* since 26 February, and *covid* since 7 March 2020.

The average daily number of tweets collected on dataset v1 was 208,662.1 (SD=100,448.7, Mdn=243,087). The number of tweets collected increased every month from 724,877 in January, 3,084,729 in February, to 2,658,920 in just the first 13 days of March. Table 1 show the summary statistics for the daily number of tweets collected each month and on the first 13 days of March.

**Table 1**. Statistics of daily tweets per month

| Month | M | SD | Mdn |
|---|---|---|---|
| January | 72,487.70 | 22964.94 | 82,153 |
| February | 106,370.00 | 40156.13 | 86,126 |
| March | 204,532.30 | 57790.74 | 163,601 |

While the dataset contains tweets from 66 languages, only English-language tweets were collected from 22 January to 30 January 2020. English-language tweets remain the most prominent in the dataset accounting for 63.4% of the total. Figure 1 presents the distribution of collected tweets by language.

**Table 2.** Distribution of Tweets per Language

| Language | Number of Tweet | Percentage |
|---|---|---|
| English | 4071165 | 63.4% |
| Spanish/Castilian | 813515 | 12.7% |
| Portuguese | 285943 | 4.5% |
| Bahasa | 272610 | 4.2% |
| French | 253950 | 4.0% |
| Italian | 246912 | 3.8% |
| Thai | 108011 | 1.7% |
| Japanese | 86871 | 1.4% |
| German | 49831 | 0.8% |
| Tagalog | 46078 | 0.7% |
| Turkish | 31260 | 0.5% |
| Catalan; Valencian | 24893 | 0.4% |
| Dutch/Flemish | 20829 | 0.3% |
| Hindi | 17832 | 0.3% |
| Chinese | 15684 | 0.2% |

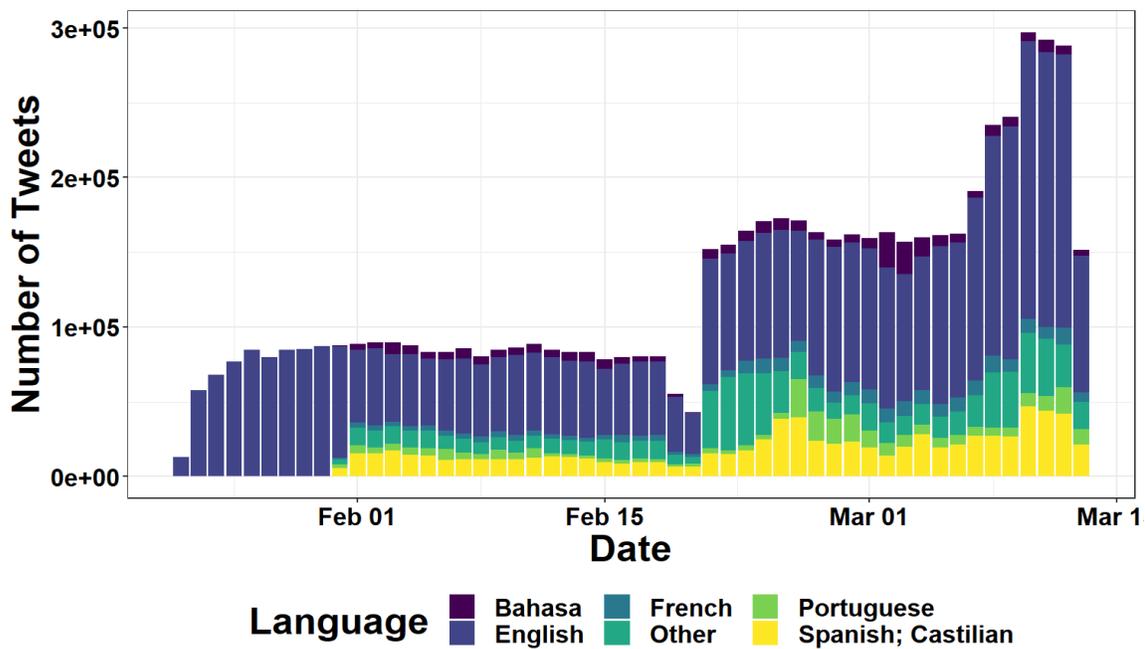

**Figure 1.** Collected tweets, by language, as of 13 March 2020.

Information about retweets and likes was also collected. Figure 2 below presents the distribution of collected tweets' retweets over the observed period. While the overall level of retweeting appears to have declined in February, retweets rose abruptly as the crisis ramped up in Europe in late February and early March.

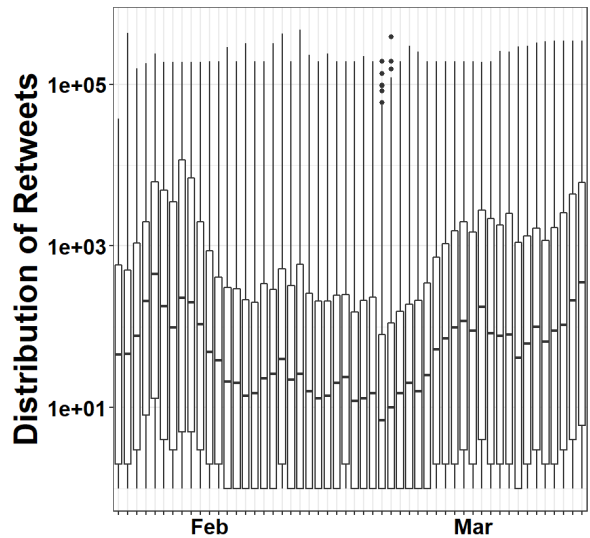

**Figure 2.** Distribution of observed retweets, on a log10 scale, across the observed period.

In addition to language, a relatively small percentage of the collected tweets contain geolocations. Figure 3 presents the locations observed as of 13 March 2020. Table 3 below provides numerical summaries of the observed tweets number of retweets, likes and geolocation information.

**Table 3** Summary statistics for collected tweets

| | |
|---|---|
| Number of Geolocated Tweets: | 1,245 |
| Maximum Observed Retweets | 469,739 |
| Median Observed Retweets: | 49 |
| 25th Percentile of Observed Retweets: | 1 |
| Maximum Observed Likes: | 653 |
| Median Observed Likes: | 0 |
| 25th Percentile of Observed Likes: | 0 |

**DATASET ACCESSIBILITY**

The dataset v1 was realized on 23 March 2020. The dataset is available on Github at : https://github.com/lopezbec/COVID19_Tweets_Dataset The dataset is released in compliance with the Twitter's Terms & Conditions. Hence, only the tweets-IDs are made available to researchers. However, using the Tweeter API the tweets can be "rehydrate" and the data of tweets that have not been deleted can be accessed (more details on the GitHub page). This dataset is still being continuously collected and routinely updated. If you have technical questions about the data collection, please contact the corresponding author.

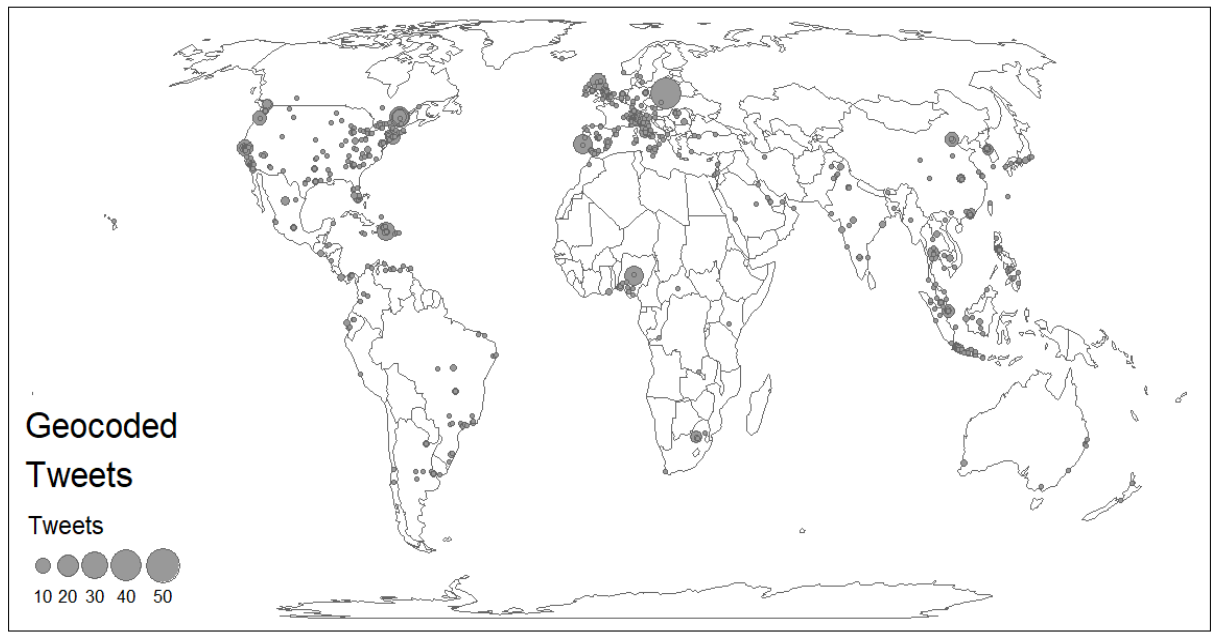

**Figure 3.** Geolocated tweets as of 13 March 2020.

**FUTURE WORK**

This research note's main objective was to introduce and share with the research community a dataset of tweets related to the COVID-19. We are continuously collecting and routinely updating the dataset. Similarity, we will be using Natural Language Processing and Text Mining, and Network Analysis to analyze the corpus of tweets to identify common popular responses to the pandemic and how these responses differ across time. Moreover, with this dataset we will explore to how information and misinformation about COVID-19 is transmitted via Twitter.